\documentstyle[preprint,aps,epsf,floats]{revtex}
\begin{document}
\tighten

%%%%%%%%%%%%%%%%%%%%%%%%%%%%%%%
%%%    private definitions  %%%
%%%%%%%%%%%%%%%%%%%%%%%%%%%%%%%
\def\bfl{{\bbox \ell}}
%\def\bfl{{\bf l}}
%%%%%%%%%%%%%
\def\bull{\vrule height .9ex width .8ex depth -.1ex}
\def\MeV{{\rm MeV}}
\def\GeV{{\rm GeV}}
\def\Tr{{\rm Tr\,}}
\def\nrcpt{NR\raise.4ex\hbox{$\chi$}PT\ }
\def\ket#1{\vert#1\rangle}
\def\bra#1{\langle#1\vert}
\def\ltap{\ \raise.3ex\hbox{$<$\kern-.75em\lower1ex\hbox{$\sim$}}\ }
\def\gtap{\ \raise.3ex\hbox{$>$\kern-.75em\lower1ex\hbox{$\sim$}}\ }
\def\abs#1{\left| #1\right|}
\def\CA{{\cal A}}
\def\CC{{\cal C}}
\def\CD{{\cal D}}
\def\CE{{\cal E}}
\def\CL{{\cal L}}
\def\CO{{\cal O}}
\def\CZ{{\cal Z}}
\def\bvert{\Bigl\vert\Bigr.}
\def\pds{{\it PDS}\ }
\def\ms{MS}
\def\ddq{{{\rm d}^dq \over (2\pi)^d}\,}
\def\ddqm{{{\rm d}^{d-1}{\bf q} \over (2\pi)^{d-1}}\,}
\def\bfq{{\bf q}}
\def\bfk{{\bf k}}
\def\bfp{{\bf p}}
\def\bfpp{{\bf p '}}
\def\bfr{{\bf r}}
\def\dtr{{\rm d}^3\bfr\,}
\def\bfx{{\bf x}}
\def\dtx{{\rm d}^3\bfx\,}
\def\dfx{{\rm d}^4 x\,}
\def\bfy{{\bf y}}
\def\dty{{\rm d}^3\bfy\,}
\def\dfy{{\rm d}^4 y\,}
\def\dfq{{{\rm d}^4 q\over (2\pi)^4}\,}
\def\dfk{{{\rm d}^4 k\over (2\pi)^4}\,}
\def\dfl{{{\rm d}^4 \ell\over (2\pi)^4}\,}
\def\dtq{{{\rm d}^3 {\bf q}\over (2\pi)^3}\,}
\def\dtk{{{\rm d}^3 {\bf k}\over (2\pi)^3}\,}
\def\dtl{{{\rm d}^3 {\bfl}\over (2\pi)^3}\,}
\def\dt{{\rm d}t\,}
\def\frac#1#2{{\textstyle{#1\over#2}}}
\def\darr#1{\raise1.5ex\hbox{$\leftrightarrow$}\mkern-16.5mu #1}
\def\){\right)}
\def\({\left( }
\def\]{\right] }
\def\[{\left[ }
\def\si{{}^1\kern-.14em S_0}
\def\siii{{}^3\kern-.14em S_1}
\def\diii{{}^3\kern-.14em D_1}
\def\dtwiii{{}^3\kern-.14em D_2}
\def\dthiii{{}^3\kern-.14em D_3}
\def\pziii{{}^3\kern-.14em P_0}
\def\poiii{{}^3\kern-.14em P_1}
\def\ptiii{{}^3\kern-.14em P_2}
\def\ipi{{}^1\kern-.14em P_1}
\def\idii{{}^1\kern-.14em D_2}
\def\fm{{\rm\ fm}}
\def\MeV{{\rm\ MeV}}
\def\CA{{\cal A}}
\def\Czzm{ {\cal A}_{-1[00]} }
\def\Cttm{{\cal A}_{-1[22]} }
\def\Ctzm{{\cal A}_{-1[20]} }
\def\Cztm{ {\cal A}_{-1[02]} }
\def\Czzz{{\cal A}_{0[00]} }
\def\Cttz{ {\cal A}_{0[22]} }
\def\Ctzz{{\cal A}_{0[20]} }
\def\Cztz{{\cal A}_{0[02]} }

\def\Ames{ A }  % The axial meson field

\newcommand{\eqn}[1]{\label{eq:#1}}
\newcommand{\refeq}[1]{(\ref{eq:#1})}
\newcommand{\eq}{eq.~\refeq}
\newcommand{\eqs}[2]{eqs.~(\ref{eq:#1}-\ref{eq:#2})}
\newcommand{\eqsii}[2]{eqs.~(\ref{eq:#1}, \ref{eq:#2})}
\newcommand{\Eq}{Eq.~\refeq}
\newcommand{\Eqs}{Eqs.~\refeq}

% A useful Journal macro
\def\Journal#1#2#3#4{{#1} {\bf #2}, #3 (#4)}

% Some useful journal names
\def\NCA{\em Nuovo Cimento}
\def\NIM{\em Nucl. Instrum. Methods}
\def\NIMA{{\em Nucl. Instrum. Methods} A}
\def\NPB{{\em Nucl. Phys.} B}
\def\NPA{{\em Nucl. Phys.} A}
\def\PLB{{\em Phys. Lett.} B}
\def\PRL{\em Phys. Rev. Lett.}
\def\PRD{{\em Phys. Rev.} D}
\def\PRC{{\em Phys. Rev.} C}
\def\PRA{{\em Phys. Rev.} A}
\def\PR{{\em Phys. Rev.} }
\def\ZPC{{\em Z. Phys.} C}
\def\SJP{{\em Sov. Phys. JETP}}
\def\SJNP{{\em Sov. Phys. Nucl. Phys.}}

\def\FBS{{\em Few Body Systems Suppl.}}
\def\IJMP{{\em Int. J. Mod. Phys.} A}
\def\UJP{{\em Ukr. J. of Phys.}}
\def\CJP{{\em Can. J. Phys.}}
\def\SCI{{\em Science} }

%%%%%% Definitions for this paper

\def\spol{\alpha_{E0}}% scalar polarizability
\def\qpol{\alpha_{E2}}% tensor polarizability
\def\Mspol{\beta_{M0}}% magnetic scalar polarizability
\def\Mqpol{\beta_{M2}}% magnetic scalar polarizability

%%%%%%%%%%%%%%%%%%%%%%%%%%%%%%%

\preprint{\vbox{
\hbox{ NT@UW-98-19}
\hbox{ CALT-68-2175}
\hbox{ DUKE-TH-98-167}
\hbox{ DOE/ER/40561-17-INT98}
}}
\bigskip
\bigskip

\title{An Effective Field Theory Calculation of 
the Parity Violating Asymmetry
in $\vec n+p\rightarrow d+\gamma$}
\author{David B. Kaplan}  
\address{ Institute for Nuclear Theory,
University of Washington, Seattle, WA 98915   
\\  {\tt dbkaplan@phys.washington.edu} }  
\author{Martin  J. Savage and Roxanne P. Springer
\footnote{ \rm  On leave from the Department of Physics, 
  Duke University, Durham NC 27708.}}  
\address{ Department of Physics, University of Washington,  
Seattle, WA 98915 
\\ {\tt savage@phys.washington.edu , rps@redhook.phys.washington.edu} }  
\author{Mark B. Wise} \address{
California Institute of Technology, Pasadena, CA 91125  
\\  {\tt wise@theory.caltech.edu} }  
\maketitle

\begin{abstract}
Weak interactions are expected to induce a parity violating pion-nucleon
coupling, $h_{\pi NN}^{(1)}$.
This coupling should be measurable in a proposed experiment to study
the parity violating asymmetry $A_\gamma$ in the process
$\vec n + p\rightarrow d+\gamma$.
We compute the leading dependence of $A_\gamma$ on the coupling
$h_{\pi NN}^{(1)}$ using recently developed effective field theory 
techniques and  find an asymmetry of  $A_\gamma = +0.17\  h_{\pi NN}^{(1)}$
at leading order.
This asymmetry has the opposite sign to that given by 
Desplanques, Donoghue and Holstein.
\end{abstract}
\vskip 1in

%\leftline{July 1998}
%%%%%%%%%%%%%%%%%%%%%%%%%%%%%%%%%
%
%        VERSION DATE
%
% \leftline{{\bf Draft version 27 July 98}}
%
%
%%%%%%%%%%%%%%%%%%%%%%%%%%%%%%%%%%
\vfill\eject

Recently an improved measurement of the parity violating asymmetry,
$A_{\gamma}$, in the angular distribution of $2.2~ {\rm MeV}$ gamma rays
from the radiative capture of  polarized cold neutrons
$\vec n+p \rightarrow d+\gamma$,
was proposed~\cite{npprop}. 
With $\theta_{{\bf s}\gamma}$ the angle between the neutron spin
and
the photon momentum, the asymmetry  $A_{\gamma}$ is defined by
\begin{equation}
{1 \over \Gamma}{d\Gamma \over d \rm{cos}\theta_{{\bf s}\gamma}}\ =\ 1\ +\
A_{\gamma}\ \rm{cos}\theta_{{\bf s}\gamma}
\ \ \  .
\end{equation}
The current experimental limit on this asymmetry parameter is
$A_\gamma = -(1.5\pm 4.8)\times 10^{-8}$~\cite{Aetal},
while the proposed  experiment expects to measure $A_{\gamma}$ with
a precision of $\pm 5 \times 10^{-9}$.
Interest in $A_{\gamma}$ is motivated by a recent 
measurement\cite{anawei} of the cesium anapole
moment that appears to give weak coupling parameters that are
inconsistent with other low energy parity violating
measurements~\cite{PVprobs,PVprobsb}.
In order to obtain the theoretically cleanest determination of 
the weak parameters in
the nucleon-meson Lagrange density it is clear that measurements in 
few nucleon systems are desirable, thereby eliminating density effects
that could arise in nuclei and are difficult to calculate.
Reviews of the the subject can be found 
in\cite{adelhax,DDH,MMa,DZ,HHpv}.

$A_\gamma$ is sensitive to the weak parity violating 
pion-nucleon coupling, 
and its measurement may provide a clean determination of
this important parameter.
In this letter we present a calculation of
$A_{\gamma}$ using recently developed effective field theory
techniques~\cite{KSW,KSW2}.
This 
method allows us to calculate processes in the two nucleon sector in a 
systematic fashion, and if carried out to higher orders, is expected to be able
to reach the same level of precision as conventional nuclear physics techniques,
but without model dependence.

The effective field theory 
method was used to calculate the electromagnetic form factors of the
deuteron in  \cite{KSW2} and the basic tools for performing the calculations in
this letter were developed there. The only new interaction needed is the parity
violation pion-nucleon coupling which appears in the interaction Lagrange
density,
\begin{equation}
{\cal L}_{pnc} =-\ {h_{\pi NN}^{(1)} \epsilon^{3ij}\over  \sqrt 2}
\ N^\dagger\ \tau^i \pi^j\ N
\ \ \ .
\label{eq:pnc}
\end{equation}
$N$ is the doublet of spin ${1 \over 2}$ nucleon fields, (i.e., $N_1=p$ and
 $N_2=n$), $\tau^i$, $i=1,2,3$ are the three Pauli matrices in isospin space and
$\pi^j$, $j=1,2,3$ are the three (real) pion fields.
The $\Delta I=1$
weak pion-nucleon coupling $h_{\pi NN}^{(1)}$ is simply related to the  
notation of Desplanques, Donoghue and Holstein (DDH)\cite{DDH},
$h_{\pi NN}^{(1)}=f_\pi$~\footnote{There are typographical errors 
  in \cite{adelhax,HHpv}.
  In eq.~3 of \cite{adelhax} and eq.~7 of \cite{HHpv} the replacement
  ${f_\pi\over 2}\rightarrow {f_\pi\over \sqrt{2}}$
  should be made. In this letter the symbol $f_\pi$ is reserved for the pion
  decay constant.
  The signs of coupling constants used in \cite{Da} are consistent with those used
  in \cite{DDH} only if the strong coupling is negative, in which case redefining
  all meson fields $M\rightarrow -M$ will give rise to the signs used
  in  \cite{DDH}.  },
but is of opposite sign to that used in \cite{KSa}.
The strong pion-nucleon interaction Lagrange density we have used is
\begin{eqnarray}
{\cal L}_{pc} & = &  {g_A\over f_\pi} 
N^\dagger {\bbox \sigma}\cdot {\bbox \nabla } \pi N
\ \ \ ,
\label{eq:pc}
\end{eqnarray}
where $\pi = \tau^i \pi^i /\sqrt{2}$, $g_A = + 1.25$ is the axial
coupling constant and $f_\pi = 132\ {\rm MeV}$ is the pion decay constant.

Predictions based on effective field theory are made in
a chiral and  momentum expansion. For cold neutron capture
$n+p\rightarrow d+\gamma$ with the neutron and proton essentially at rest
the relevant momentum $Q$ is determined by the binding energy of the deuteron,
$B=2.224~{\rm MeV}$ and is $\sqrt{M_N B} = 45.70\ {\rm MeV}$, where $M_N$
is the nucleon mass.
The power counting treats $Q/\Lambda_{QCD}$ and
$m_{\pi}/\Lambda_{QCD}$ 
(where $\Lambda_{QCD}$ is the scale characteristic of short range 
nucleon-nucleon interactions)
as small and takes $Q \sim m_{\pi}$ . This is similar
in spirit to applications of chiral perturbation theory in the single nucleon
sector and for pion self interactions.
However, the power counting is
unlike that used in conventional chiral perturbation theory
because of the large
scattering lengths that occur in the $^1S_0$ and $^3S_1$ $NN$ scattering channels. 
These large scattering lengths render the leading order interactions
nonperturbative
and cause the two-body operators to develop large anomalous dimensions.
The expansion is described in detail in Refs.~\cite{KSW,KSW2,SSp}.
At lowest order in the $Q$ power counting ${\cal L}_{pnc}$, eq.~(\ref{eq:pnc}), 
is the only 
parity violating interaction that occurs~\cite{KSa,SSp}. 
Other terms, such as the parity violating two-body operators are not 
relevant until higher order in the $Q$ expansion~\cite{SSp}.

At leading order in the $Q$ expansion we find the matrix element for
cold neutron capture $n+p\rightarrow d+\gamma$
can be written as
\begin{eqnarray}
\label{eq:matrix}
{\cal M} & = &
e\ X\ N^T\tau_2\ {\bbox \sigma_2} \  \left[ {\bbox \sigma}\cdot {\bf q}\ 
\ {\bbox\epsilon} (d)^* \cdot {\bbox \epsilon} (\gamma)^*
  \ -\ {\bbox \sigma} \cdot  {\bbox \epsilon} (\gamma)^*\ 
  \ {\bf q}\cdot {\bbox \epsilon} (d)^* 
  \right] N 
\\ \nonumber
& + &
i e\ Y\  \epsilon^{ijk}\ {\bbox \epsilon} (d)^{i*}\   
{\bf q}^j\  {\bbox\epsilon} (\gamma)^{k*}
\ (N^T\tau^2 \tau^3 {\bbox\sigma}^2 N)
\\ \nonumber
& + & i e\ W\ \epsilon^{ijk}\ {\bbox\epsilon} (d)^{i*}\  
{\bbox\epsilon}(\gamma)^{k*}\
(N^T\tau^2 {\bbox \sigma^2}{\bbox \sigma^j} N)
\ \ \ \ .
\end{eqnarray}
In eq.~(\ref{eq:matrix})  $e=|e|$ is the magnitude of the  electron charge, 
$N$ is 
the doublet of nucleon spinors, ${\bbox \epsilon}(\gamma)$ is 
the polarization vector for the photon, ${\bbox \epsilon}(d)$ is the polarization
vector for the deuteron and ${\bf q}$ is the outgoing photon momentum.
The terms $X$ and $Y$ are parity conserving while the term $W$ is parity
violating.
Note that for the parity conserving term $Y$ the neutron and proton are
in a $^1S_0$ state while for the parity conserving term $X$ and the
parity violating term $W$ they are in a $^3S_1$
state.
Interference between the parity conserving and parity violating 
amplitudes is possible if the neutron is polarized.
%%%%%%%%%%   Figure 1   %%%%%%%%%%%%%%%
\begin{figure}[t]
\centerline{{\epsfxsize=3.0in \epsfbox{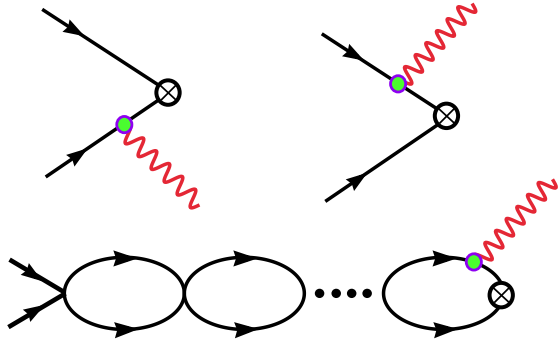}} }
\noindent
\caption{\it Graphs contributing to the parity conserving amplitude  
  for $n+p\rightarrow d+\gamma$ at leading order in the
  effective field theory expansion. 
  The solid lines denote nucleons
  and the wavy lines denote photons.
The light solid circles correspond to the nucleon magnetic
moment coupling to
the electromagnetic field.
  The crossed circle represents an insertion of the deuteron
  interpolating 
  field which is taken to have $\siii$ quantum numbers.
}
\label{fig:strong}
\vskip .2in
\end{figure}
At leading order,
$X$ and $Y$ are calculated from the sum of Feynman diagrams 
in Fig.~(\ref{fig:strong})
and from wavefunction renormalization associated with the deuteron
interpolating
field\cite{KSW2}, giving  
\begin{equation}
  X\ =\ 0\ \ \ \  ,\ \ \ \ 
Y\ =\ -\sqrt{\pi\over\gamma^3}\ { 2\kappa_1  \over M_N}
\ \left( 1-\gamma a_0 \right)
\ \ \ ,
\end{equation}
where $\kappa_1 ={1\over 2}(\kappa_p-\kappa_n)= 2.35294$ is the 
isovector nucleon magnetic moment in units of nuclear magnetons,
$a_0=-23.714 \pm 0.013~  {\rm fm}$ is the $NN$ $^1S_0$ scattering length,
and $\gamma=\sqrt {M_N B}$.
This expression for $Y$ yields the
$n+p \rightarrow d+ \gamma$ capture cross section
\begin{equation}
\label{eq:leading}
\sigma={8\pi\alpha\gamma^5\kappa_1^2 a_0^2 \over v M_N^5}
\left(1-{1 \over \gamma a_0} \right)^2, 
\end{equation}
where $\alpha$ is the fine structure constant and $v$ is the magnitude of the
neutron velocity (in the proton rest frame).
This agrees with the results of Bethe and Longmire~\cite{BLa,Noyes}
when terms in their expression involving the effective range are neglected.
Eq.~(\ref{eq:leading}) is about $10\%$ less than the experimental value for the
cross section.
In the power counting appropriate to the effective
field theory approach the effective range enters at next order
in the $Q$ counting. However, other effects also occur at this order.
For example, a two body operator involving the magnetic field.
Including just the effective range does not represent a systematic
improvement of the theoretical expression in eq.~(\ref{eq:leading}).
This strong interaction amplitude was also examined in ref.~\cite{PMRa}
using effective field theory methods.

\begin{figure}[t]
\centerline{{\epsfxsize=4.0in \epsfbox{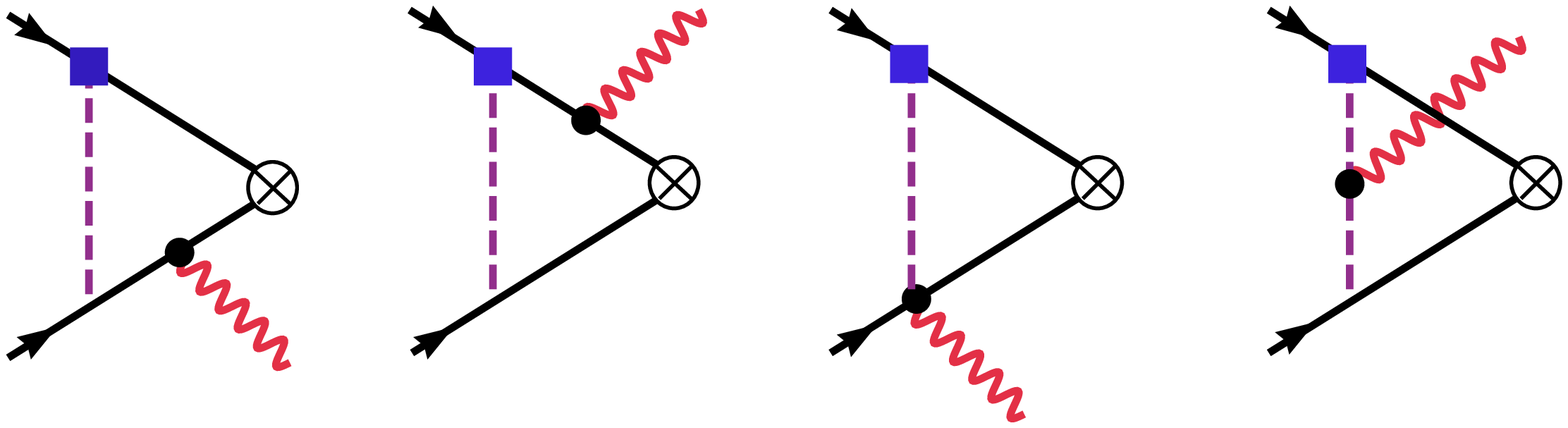}} }
\vskip 0.6in
\centerline{{\epsfxsize=4.0in \epsfbox{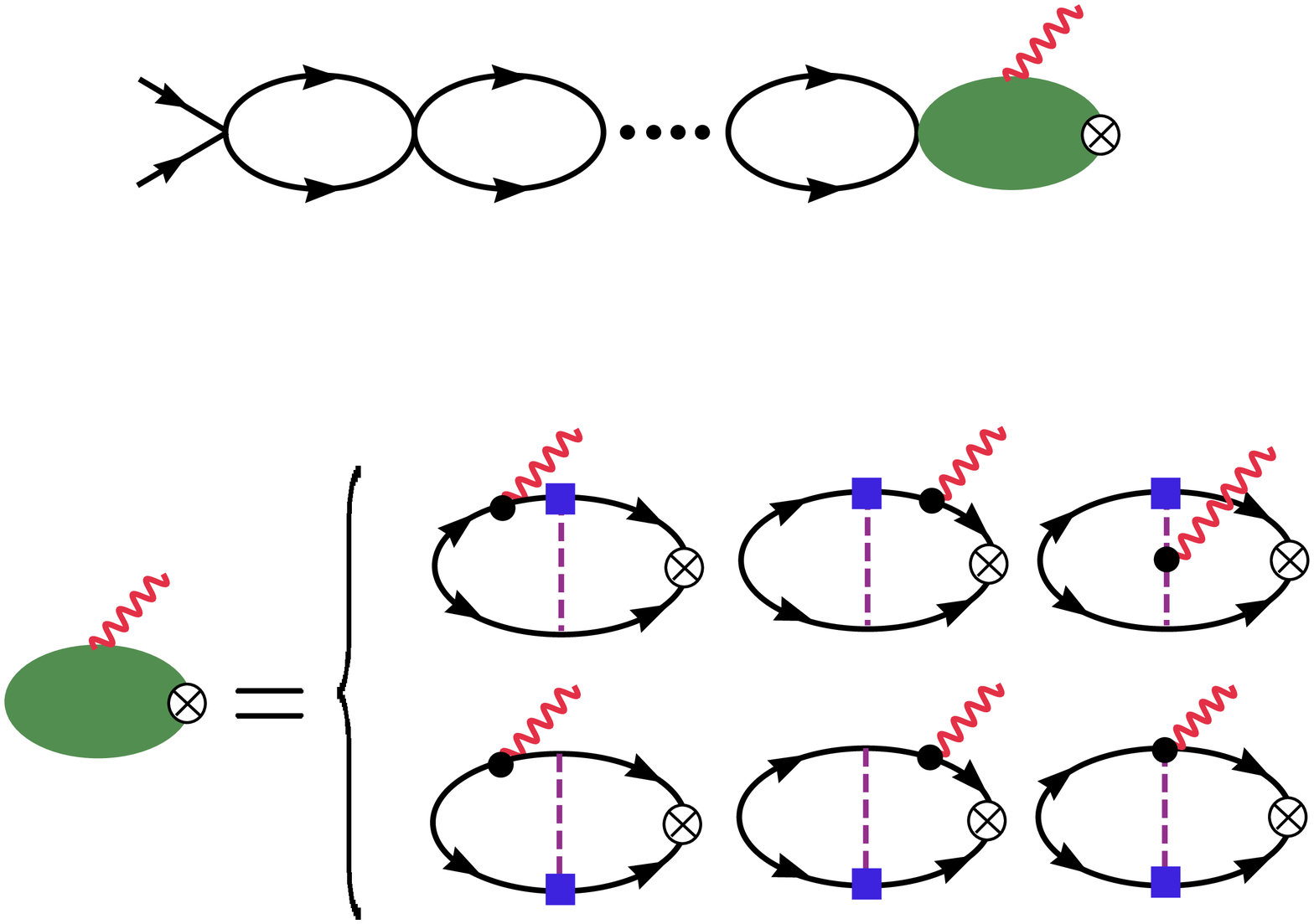}} }
\noindent
\caption{\it Graphs contributing to the parity violating amplitude
  for $n+p\rightarrow d+\gamma$ at leading order in the
  effective field theory
  expansion.
  The solid lines denote nucleons, the dashed lines denote
  pions and the wavy
  lines denote photons.
  The solid squares denote an  insertion of ${\cal L}_{pnc}$ 
  (eq.~(\ref{eq:pnc}) ) while
  the solid circles correspond to the minimal electromagnetic coupling.
  The crossed circle represents an insertion of the deuteron interpolating 
  field which is taken to have $\siii$ quantum numbers.
}
\label{fig:weak}
\vskip .2in
\end{figure}

In terms of the amplitudes $X$, $Y$ and  $W$ the parity violating asymmetry is,
\begin{equation}
A_\gamma\ =\ -{2 M_N\over \gamma^2}\  {{\rm Re}[(Y+X)^*\ W]\over 2|X|^2 + |Y|^2}
\ \ \ \  ,
\end{equation}
where $\gamma^2/M_N$ is the photon energy.
At leading order in the $Q$ expansion $W$ follows from the sum of diagrams
in Fig.~(\ref{fig:weak}).
%%%%%%%%%%   Figure 2   %%%%%%%%%%%%%%%
We find that
\begin{eqnarray}
W & = & g_A\  h_{\pi NN}^{(1)}\ {\sqrt{\pi\gamma} \over 3\pi f_{\pi}}
\left[ {m_{\pi} \over (m{_\pi}+\gamma)^2}
-{m_{\pi}^2 \over 2 \gamma^3}{\rm ln }\left({2\gamma \over m_{\pi}}+1 \right)+
{m_{\pi}^2 \over \gamma^2 (m_{\pi}+\gamma)}\right]
\ \ \  ,
\end{eqnarray}
where $g_A$ is defined in eq.~(\ref{eq:pc}),
$h_{\pi NN}^{(1)}$ is defined in  eq.~(\ref{eq:pnc}),
and $m_{\pi} \simeq 140~{\rm MeV}$ is the pion mass.
The first term in the square brackets comes from the
one loop diagrams while the second and third terms come from the two loop
diagrams with the neutron and proton rescattering through a contact term.
Therefore, at leading order in the effective field
theory $Q$ expansion the numerical value of the asymmetry $A_\gamma$ is,
\begin{eqnarray}
  A_\gamma & = & +0.17\  h_{\pi NN}^{(1)}
  \ \ \ .
\label{eq:asymm}
\end{eqnarray}
A naive dimensional analysis estimate of the weak coupling\cite{KSa} yields
$|h_{\pi NN}^{(1)}|\sim 5\times 10^{-7}$, arising largely from the strange quark
operators\cite{DSLS}, consistent with the best guess of DDH
(a recent calculation in the $SU(3)$ Skyrme model yields 
$h_{\pi NN}^{(1)}\sim +1.3\times 10^{-7}$\cite{UGMW}).
Hence an asymmetry $|A_\gamma|\sim 0.8\times 10^{-7}$
could reasonably be expected.
This is consistent with the present experimental
bound and would be easily accessible to the experiment proposed 
in Ref.~\cite{npprop}.

The calculated asymmetry in eq.~(\ref{eq:asymm})
is somewhat larger in magnitude than  previous calculations that have found
$ A_\gamma  \sim  -0.11\  h_{\pi NN}^{(1)}$~\cite{Dani,Tad,Da,DMa,Bruce}.
It is also of the opposite sign to the currently accepted theoretical
prediction~\cite{DMa}.
However, as stated in a footnote in \cite{DMa},
the sign of the asymmetry computed in
\cite{DMa} disagrees with the sign computed in \cite{Dani,Bruce}.

The Feynman diagrams that contribute to $W$ contain a contribution from
exchange currents where the photon couples to the exchanged pion.
This contribution by itself is ultraviolet divergent, yet the sum of diagrams
is finite. This means that the value of the exchange 
current contribution alone is
dependent on the ultraviolet regulator and subtraction scheme adopted.
In potential models, the short distance behavior of the potential regulates the
ultraviolet behavior. Many different models for the short distance behavior
of the potential give the same low energy physics and the size of the
exchange current contribution to $W$ depends on how the short distance physics is modeled.
For the parity conserving amplitude $Y$ the exchange contribution does not
occur until next-to-leading order in the $Q$ expansion. 
A similar situation occurs there
and again the value of the exchange current contribution alone 
is not a meaningful quantity
since in the effective field theory approach that contribution is ultraviolet
divergent and its value depends on the regulator and subtraction scheme used.

In the $Q$ expansion the leading contribution to $W$ 
is $\sim 1/\sqrt{Q}$. At next-to-leading
order (i.e., $\sim \sqrt{Q}$) the parity violating $S$-wave to
$P$-wave two body operators contribute. Since their coefficients are not known
it is not possible at this time to improve the calculation of $W$ 
by going beyond leading order in the $Q$ expansion.
However, the same coefficients appear in other parity violating observables
so that a systematic analysis of higher order effects may be possible.

To conclude, we have computed the parity violating symmetry $A_\gamma$
in the radiative capture of polarized cold neutrons
$\vec n + p\rightarrow d+\gamma$ at leading order in effective field
theory.
The asymmetry $A_\gamma$ will provide a relatively
clean determination of $h_{\pi NN}^{(1)}$, 
unless this coupling is anomalously small, as suggested by
the circular polarization experiments in $^{18}F$\cite{Fl}.
If it is much smaller than naive estimates suggest then there
will be additional contributions from parity violating two body operators that
would need to be included.

\vskip 0.5in

We would like to thank Wick Haxton and Bruce McKellar for useful discussions.
This work is supported in part by the U.S. Dept. of Energy under
Grants No. DE-FG03-97ER4014 and DE-FG02-96ER40945.

\end{document}